\documentclass[conference,10pt]{IEEEtran}
\normalsize
\ifCLASSINFOpdf
\else
\fi

\usepackage{amsthm}
\usepackage{amsmath}
\usepackage{amssymb}
\usepackage{graphicx}
\usepackage{esint}
\usepackage{amsfonts}
\usepackage{cite}
\usepackage{balance}
\usepackage{caption}
\usepackage{subcaption}
\usepackage{epstopdf}
\usepackage{color}
\usepackage{tikz}

\makeatletter

\theoremstyle{plain}

\makeatother

\providecommand{\theoremname}{Theorem}

\theoremstyle{plain}

\makeatother

\providecommand{\lemmaname}{Lemma}

\theoremstyle{plain}

\makeatother

\providecommand{\corname}{Corollary}

\theoremstyle{plain}
\newtheorem{prop}{\protect\propname}

\makeatother

\providecommand{\propname}{Proposition}

\DeclareMathOperator{\erf}{erf}

\usepackage{pifont}

\newcommand\blfootnote[1]{%
	\begingroup
	\renewcommand\thefootnote{}\footnote{#1}%
	\addtocounter{footnote}{-1}%
	\endgroup
}

\IEEEoverridecommandlockouts\IEEEpubid{\makebox[\columnwidth]{ 978-1-6654-3540-6/22~\copyright~2022 IEEE \hfill} \hspace{\columnsep}\makebox[\columnwidth]{ }}

\begin{document}
\title{Throughput analysis of RIS-assisted UAV wireless systems under disorientation and misalignment}

\author{
	Alexandros-Apostolos A. Boulogeorgos$\,^{1}$, Angeliki Alexiou$\,^{1}$, and   Marco Di Renzo$\,^{2}$  
	\\
	\footnotesize{$\,^{1}$Digital Systems, University of Piraeus Piraeus 18534 Greece.} e-mails: al.boulogeorgos@ieee.org, alexiou@unipi.gr.\\
	$\,^{2}$Universit\'e Paris-Saclay, CNRS, CentraleSup\'elec, Laboratoire des Signaux et Syst\`emes, 3 Rue Joliot-Curie, 91192 Gif-sur-Yvette, France. \\
	e-mail: marco.di-renzo@universite-paris-saclay.fr
	
}
\maketitle	
\begin{abstract}
  Reconfigurable intelligent surface (RIS)-assisted unmanned areal vehicles (UAV)  communications have been identified as a key enabler of a number of next-generation applications. However, to the best of our knowledge, there is no generalized framework for the quantification of the throughput performance of RIS-assisted UAV systems. Motivated by this, in this paper, we present a comprehensive system model that accounts the impact of multi-path fading, which is modeled by means of mixture gamma, transceiver hardware imperfections, and stochastic beam disorientation and misalignment in order to examine the throughput performance of a RIS-assisted UAV wireless system.  In this direction, we present a novel closed-form expression for the systems throughput for two scenarios: i) in the presence and ii) in the absence of disorientation and misalignment. Interestingly, our results reveal the importance of accurate modeling the aforementioned phenomena as well as the existence of an optimal transmission spectral efficiency.\blfootnote{This paper is the conference version of~\cite{Boulogeorgos2022}, which have been submitted form possible publication in IEEE Transactions of Vehicular Technology.}\blfootnote{This work has received funding from the European Commission Horizon 2020 research and innovation programme ARIADNE under grant agreement No. 871464.}    
\end{abstract}

\section{Introduction}\label{S:Intro}

Unmanned aerial vehicles (UAVs) have been widely recognized as one of the key enablers of beyond the fifth generation (B5G) networks~\cite{Pliatsios2021,Geraci2021}. 
From communications point of view, all UAV-based system models have two common requirements: i) ultra-reliable connectivity ~\cite{Xiao2021,Boulogeorgos2021a}, and ii) high-energy efficiency~\cite{Zeng2017,Babu2021,PhD:Boulogeorgos}. These requirements can be technically translated into the necessity of creating a beneficial and reconfigurable electromagnetic environment without decreasing the UAV's energy efficiency. 

Recognizing the aforementioned need, several researchers have recently focused their attention on combining UAVs and reconfigurable intelligent surfaces (RIS)~\cite{Agrawal2021,Guo2021,Pan2021,Yang2020}.  In more detail, in~\cite{Agrawal2021}, the performance of an interference-limited RIS-assisted UAV wireless system suffering from Nakagami-$m$ fading was evaluated in terms of coverage probability, bit and block error rates.
   In~\cite{Guo2021}, the active beamforming of the UAV, the coefficients of the RIS MAs, and the trajectory of the UAV were jointly optimized to maximize the overall secrecy rate of all legitimate users in the presence of multiple eavesdroppers in RIS-assisted UAV wireless systems operating in the millimeter-wave band and exhibiting Rayleigh fading. In~\cite{Pan2021}, the authors investigated the joint optimization of UAV trajectory, RIS PS, sub-band allocation, and  power management  to maximize the minimum average achievable rate of all users in a RIS-assisted UAV wireless system operating in the terahertz (THz) band. Finally, in~\cite{Yang2020}, the OP, average bit error rate, and ergodic capacity of RIS-assisted dual-hop UAV communication systems were quantified, where the S-RIS and RIS-UAV links were  modeled as Rayleigh and mixture-Gamma (MG) RVs, respectively.

To the best of the authors' knowledge, no generalized channel model has yet been presented that captures the specifics of RIS-assisted UAV wireless systems and enables performance evaluation in different propagation environments.    
Likewise, most of the aforementioned contributions assume that the RIS-UAV link is not directional. As a result, the detrimental effect of UAV disorientation and/or  misalignment of the RIS-UAV beam have been neglected. However, as we move toward higher operating frequency, the directionality of links is expected to increase~\cite{Boulogeorgos2021a,Boulogeorgos2021,Boulogeorgos2018,C:UserAssociationInUltraDenseTHzNetworks}. Hence, even small disorientations and/or misalignments may adversely affect the performance of the RIS-assisted UAV wireless system. Aside from the joint effect of disorientation and misalignment, another important performance-limiting factor in high-frequency communications is the effect of transceiver hardware imperfections~\cite{B:Schenk-book,A:IQSC,A:Effects_of_RF_impairments_In_Cascaded}. 

Motivated by this, this work derives a generalized  framework for evaluating the throughput performance of RIS-assisted UAV wireless systems that captures the effects of various fading conditions, disorientation, misalignment, and/or hardware imperfections. The technical contribution lies in the presentation of a comprehensive system model that accounts for the effects of various small-scale fading conditions, the joint effect of stochastic disorientation and misalignment, as well as transceiver hardware imperfections. In contrast to previous publications, we assume independent MG fading in both S-RIS and RIS-UAV channels. Moreover, we closed-form expressions, simplified for the following cases: (i)  the RIS-UAV link experience neither disorientation nor misalignment, (ii) the RIS-UAV link experience disorientation and misalignment.

\subsubsection*{Notations} 
Unless otherwise stated, lower bold letter stands for vectors.  
 $\Pr\left(\mathcal{A}\right)$ denotes the probability for the event $\mathcal{A}$ to be valid. Likewise, $\cos\left(x\right)$ gives the cosine of $x$, while $\sin\left(x\right)$ returns the sine of $x$. Moreover, $\csc\left(x\right)$ stands for the cosecant of $x$. The error-function is represented by $\erf\left(\cdot\right)$~\cite[eq. (8.250/1)]{B:Gra_Ryz_Book}.
The modified Bessel function of the second kind of order $n$ is denoted as~$\mathrm{K}_n(\cdot)$~\cite[Eq. (8.407/1)]{B:Gra_Ryz_Book}. 
The  Gamma~\cite[Eq. (8.310)]{B:Gra_Ryz_Book} function is  denoted by  $\Gamma\left(\cdot\right)$. Finally, $\,_pF_q\left(a_1, a_2, \cdots, a_p; b_1, b_2, \cdots, b_q; x\right)$ and $G_{p, q}^{m, n}\left(x\left| \begin{array}{c} a_1, a_2, \cdots, a_{p} \\ b_{1}, b_2, \cdots, b_q\end{array}\right.\right)$  respectively stand for the generalized hypergeometric function~\cite[Eq. (9.111)]{B:Gra_Ryz_Book} and the Meijer G-function~\cite[Eq. (9.301)]{B:Gra_Ryz_Book}.

\section{System model}\label{S:SM}\vspace{-0.2cm}

\begin{figure}
	\centering
	\scalebox{0.44}{\input{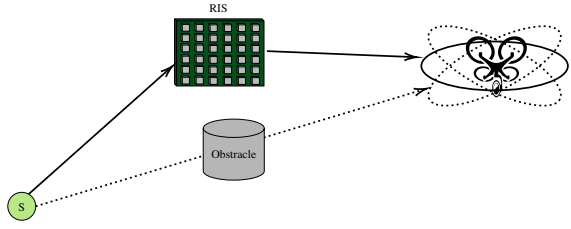}}
	\caption{System model}
	\label{Fig:SM}
	\vspace{-0.2cm}
\end{figure}

\begin{figure}
	\centering
	\scalebox{0.5}{\input{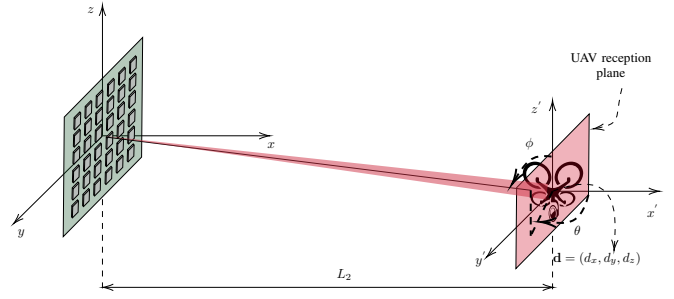}}
	\vspace{-0.2cm}
	\caption{The relative position and orientation of the RIS and UAV in the considered coordinate systems.}
	\label{Fig:Coordinates}
	\vspace{-0.4cm}
\end{figure}

As shown in Fig.~\ref{Fig:SM}, we consider an RIS-assisted UAV wireless system in which the S communicates with a UAV via a RIS. We assume that no direct link can be established between S and the UAV, due to the presence of blockage. It is assumed that the UAV is in a hover state, where, in practice, both the position and orientation of the UAV are not completely fixed.  It is also assumed that both the S and the UAV are equipped with single antennas, while the RIS consists of $N$ MAs. 
To determine the relative position and orientation of the RIS and the UAV, we define two three-dimensional Cartesian coordinate systems, as demonstrated in Fig.~\ref{Fig:Coordinates}. The RIS is assumed to be at the origin of the Cartesian coordinate system $(x, y, z)$, i.e., at position $(0,0,0)$. At a specific timeslot, it is assumed that the UAV is located at position
	$\mathbf{d}_{\epsilon}=\left(d_x+\epsilon_x, d_y+\epsilon_y, d_z+\epsilon_z\right)$,
	with respect to the $(x, y, z)$ coordinate system, where 
		$\mathbf{d}=\left(d_x, d_y, d_z\right)$
	denotes the mean of the random vector $\mathbf{d}_{\epsilon}$
	and 
	$\boldsymbol{\epsilon}=\left(\epsilon_x,\epsilon_y,\epsilon_z\right)$ 
	are independent and identical zero-mean Gaussian distributed RVs with variance $\sigma_p^2$. For a given $\mathbf{d}_{\epsilon}$, a second Cartesian coordinate system $(x^{'}, y^{'}, z^{'})$, for which $\mathbf{d}_{\epsilon}$ is at the origin and the axes $x^{'}$, $y^{'}$, and $z^{'}$ are parallel with the axes $x$, $y$, and $z$, respectively, is considered. Let $\theta_\epsilon\in[0, 2\pi]$ be the angle between the axis $x^{'}$ and the projection of the beam vector onto the $x^{'}-y^{'}$ plane. Similarly,  $\phi_{\epsilon}\in[0,\pi]$ represents the angle between the $z^{'}$ axis and the beam vector. Notice that both $\theta_\epsilon$ and $\phi_{\epsilon}$ are randomly distributed variables that can be respectively expressed~as
	$\theta_\epsilon = \theta + \epsilon_{\theta}$
and 
	$\phi_\epsilon = \phi + \epsilon_{\phi}$,
where $\theta$ and $\phi$  denote the mean of $\theta_\epsilon$ and $\phi_\epsilon$, respectively, while $\epsilon_{\theta}$ and $\epsilon_{\phi}$ are zero-mean RVs with variance $\sigma_o^2$.

By $h_i$ and $g_i$ we denote the complex coefficients of the S to $i-$th MA and $i-$th MA to UAV baseband equivalent fading channels, respectively. It is assumed that all the fading channels considered are independent, identical, slowly varying, and flat. Finally, we use $\theta_i$ to denote the response of the $i-$th MA. As a result, the baseband equivalent received signal at D can be obtained~as
\begin{align}
	r \hspace{-0.15cm} = \hspace{-0.15cm} \left\{\hspace{-0.2cm}\begin{array}{l l} 
		h_l\, A\,  \left( s+ \eta_s\right) + \eta_d + w, & \hspace{-0.25cm} \text{without disorientation} \\ & \hspace{-0.25cm} \text{or misalignment}\\
		h_l\, h_g\, A\,  \left( s+ \eta_s\right) + \eta_d + w, & \hspace{-0.25cm} \text{with disorientation} \\ & \hspace{-0.25cm} \text{and misalignment}\end{array}\right.
	\label{Eq:r}
\end{align}
where  $s$ represents the S transmission symbol, while $w$ stands for the additive white Gaussian noise (AWGN), which is modeled as a zero-mean complex Gaussian (ZMCG) RV with variance equal to $\sigma_w^2$. Moreover, $h_l$ denotes the e2e spreading-loss coefficient and can be written~as \vspace{-0.2cm}
	$h_l = h_{l,1}\,h_{l,2},$
where 
\begin{align}
	h_{l,i} = l_i\,L_{i}^{-n_i/2},\text{ with } i\in\{1, 2\},
	\label{Eq:h_l_i}
\end{align}
is the spreading loss coefficient of the S-RIS ($i=1$) and RIS-UAV ($i=2$) links respectively. In~\eqref{Eq:h_l_i}, $l_1$ and $l_2$ respectively denote the square root of the reference reception power, while $n_1$ and $n_2$ stand for the path-loss exponent of the S-RIS and RIS-UAV links, respectively. The transmission distances of the S-RIS and RIS-D links are denoted by $L_1$ and $L_2$, respectively.   

In addition, $h_g$ represents the geometric-loss caused by the disorientation of the UAV and the misalignment of the beam between the RIS and the UAV. According to~\cite{Najafi2018}, the PDF of $h_g$ can be expressed~as\vspace{-0.3cm}
\begin{align}
	f_{h_g}(x) = \frac{\rho}{B_o} \left(\frac{x}{B_o}\right)^{\zeta-1}, \text{ with } 0\leq x \leq B_o.
	\label{Eq:f_h_g}
\end{align}
In~\eqref{Eq:f_h_g}, 
	$B_o = \erf\left(v_{\min}\right) \erf\left(v_{\max}\right),$
where 
\begin{align}
	v_{i} = \frac{\alpha}{w(L_2)} \sqrt{\frac{\pi}{2\rho_{i}}},\text{ with } i\in\{\min, \max\}.
	\label{Eq:v_i}
\end{align}
In~\eqref{Eq:v_i}, 
	$\rho_{\min} = \frac{2}{\rho_y+\rho_z + \sqrt{\left(\rho_y-\rho_z\right)^2 + 4 \rho_{zy}^2}}$
and
	$\rho_{\max} = \frac{2}{\rho_y+\rho_z - \sqrt{\left(\rho_y-\rho_z\right)^2 + 4 \rho_{zy}^2}}$
where 
	$\rho_y = \cos^2\left(\phi\right) + \sin^2\left(\phi\right) \cos^2\left(\theta\right)$, 
	$\rho_z = \sin^2\left(\phi\right)$
and
$	\rho_{yz} = - \cos\left(\phi\right) \sin\left(\phi\right) \sin\left(\theta\right).$	
Moreover, $\alpha$ is the radius of the UAV antenna effective area, while $w(L_2)$ is  the RIS beamwidth at distance $L_2$ and can be expressed~as
	$w\left(L_2\right) = w_o \sqrt{1+\left(1+\frac{2 w_o^2}{\rho^2\left(L_2\right)} \right)\left( \frac{c L_2}{\pi f w_o^2} \right)^2},$ 
with $w_o$ being the beam-waist radius, $c$ denoting the speed of light, $f$ standing for the transmission frequency, and $\rho\left(L_2\right)$ representing the coherence length. Of note, the coherence length can be evaluated~as
	$\rho\left(L_2\right) = \left(0.55\, C_n^2\, k^2 \, L_2\right)^{-3/5},$
where $C_n^2$ stands for the index of refraction structure parameter and 
	$k=\frac{2\pi f}{c},$
	is the wave-number.  
In~\eqref{Eq:f_h_g},
	$\zeta = \frac{k_m \left(w(L_2)\right)^2}{4\sigma_p^2+4 d_x^2 \sigma_0^2}.$
with
	$k_m = \frac{k_{\min}+k_{\max}}{2},$
where 
	$k_{i}=\frac{\sqrt{\pi}\rho_{i} \erf\left(v_{i}\right)}{2v_{i}\exp\left(-v_{i}^2\right)}, \text{ with } i\in\left\{\min, \max\right\}.$

Finally, 
\begin{align}
	A= \sum_{i=1}^{N} h_i \theta_i g_i, 
	\label{Eq:A}
\end{align}     
where $h_i$ and $g_i$ are the S-$i-$th MA and $i-$th MA-D channel coefficients, which can be expressed as 
\begin{align}
	h_i = \left|h_i\right| \exp(j\phi_{h_i})
\text{ and }
	g_i = \left|g_i\right| \exp(j\phi_{g_i}),
\end{align}
with $\phi_{h_i}$ and $\phi_{g_i}$ respectively being the phases of $h_i$ and $g_i$. The envelops of $h_i$ and $g_i$ are assumed to be independent MG random variables with PDFs that can be respectively expressed~as 
\begin{align}
	f_{h_i}(x) = \sum_{m=1}^{M} 2 a^{(1)}_{m} x^{2 b^{(1)}_m-1}\exp\left(-c_1 x^2\right)
	\label{Eq:f_h_i}
\end{align}
and \vspace{-0.2cm}
\begin{align}
	f_{g_i}(x) = \sum_{k=1}^{K} 2 a^{(2)}_{k} x^{2 b^{(2)}_k-1}\exp\left(-c_2 x^2\right),
	\label{Eq:f_g_i}
\end{align} 
 where $M$ and $K$ are the numbers of terms for the approximation of the PDF of $\left|h_i\right|$ and $\left|g_i\right|$, respectively, while $a^{(1)}_{m}$, $b^{(1)}_m$ and $c_1$ with $m\in[1, M]$ are parameters of the $m-$th term of~\eqref{Eq:f_h_i}. Similarly, $a^{(2)}_{k}$, $b^{(2)}_k$ and $c_2$ with $k\in[1, K]$ are parameters of the $k-$th term of~\eqref{Eq:f_g_i}.

Moreover, in~\eqref{Eq:A}, $\theta_i$ stands for the $i-$th MA response and can be further written~as 
\begin{align}
	\theta_i = \left|\theta_i\right| \exp\left(j\phi_i\right), 
	\label{Eq:theta_i} 
\end{align} 
with $\left|\theta_i\right|$ and $\phi_i$ respectively representing the $i-$th MA response gain and the PS applied by the $i-$th MA of the RIS. Without loss of generality, we assume that $\left|\theta_i\right|=1$. Based on~\cite{Asadchy2016}, this is considered a realistic assumption. As reported in several works including~\cite{Boulogeorgos2020a,Basar2019}, the optimal PS for the $i-$th MA~is  \vspace{-0.3cm}
\begin{align}
	\phi_i = -\phi_{h_i} - \phi_{g_i}. 
	\label{Eq:phi_i_optimal}
\end{align} 
By applying~\eqref{Eq:phi_i_optimal} to~\eqref{Eq:A}, we get  \vspace{-0.05cm}
\begin{align}
	A=\sum_{i=1}^{N} \left|h_i\right| \left|g_i\right|. 
	\label{Eq:A_final}
\end{align}

In~\eqref{Eq:r}, $\eta_s$ and $\eta_d$ models respectively for the impact of S and UAV RF chains hardware imperfections. According to~\cite{Boulogeorgos2016,C:Energy_Detection_under_RF_impairments_for_CR}, for a given e2e channel realization,  $\eta_s$ and $\eta_d$ can be modeled as two independent ZMCG RVs with variances that can be respectively expressed~as  
	$\sigma_s^2=\kappa_t^2\,P_s$
and  
	$\sigma_d^2 = \kappa_r^2\, h_l^2\, h_g^2\, A^2\, P_s,$
where $\kappa_t$ and $\kappa_r$ are respectively the S transmitter and UAV receiver error vector magnitudes, while $P_s$ stands for the average transmitted power. According to~\cite{Boulogeorgos2019,Boulogeorgos2020}, $\kappa_t$ and $\kappa_r$ in high-frequency systems, such as millimeter wave and terahertz (THz), are in the range of $[0.07, 0.3]$. Finally, for the special case in which both the S and UAV are equipped with ideal transceivers, $\kappa_t=\kappa_r=0$~\cite{Bjoernson2013}.  

\section{Throughput Analysis}\label{S:E2eChannel} 
The signal-to-distortion-plus-noise-ratio (SDNR) at the UAV can be expressed~as  
\begin{align}
	\gamma_u\hspace{-0.1cm} =\hspace{-0.1cm} \left\{ \hspace{-0.1cm}
	\begin{array}{c l}
		\frac{h_l^2 A^2 P_s}{\left(\kappa_t^2+\kappa_r^2\right) h_l^2 A^2 P_s + \sigma_w^2}, & \text{ without disorientation}  \vspace{-0.2cm} \\ \vspace{-0.2cm}  &  \text{ or misalignment}\\
		\frac{h_l^2 A_{e2e}^2 P_s}{\left(\kappa_t^2+\kappa_r^2\right) h_l^2 A_{e2e}^2 P_s + \sigma_w^2}, & \text{ with disorientation}  \vspace{-0.2cm} \\  &  \vspace{-0.2cm} \text{ and misalignment}
	\end{array}\right.
\end{align}
or equivalently \vspace{-0.2cm}
\begin{align}
	\gamma_u\hspace{-0.1cm} =\hspace{-0.1cm} \left\{ 
	\begin{array}{c l}
		\frac{A^2}{\left(\kappa_t^2+\kappa_r^2\right) A^2 + \frac{1}{\gamma}}, & \text{ without disorientation} \\  &  \text{ or misalignment}\\
		\frac{A_{e2e}^2}{\left(\kappa_t^2+\kappa_r^2\right) A_{e2e}^2 + \frac{1}{\gamma}}, & \text{ with disorientation} \\  &  \text{ and misalignment}
	\end{array}\right..
	\label{Eq:gamma_u}
\end{align}

\subsection{Without disorientation and misalignment}\label{SS:OP_WHI} 
 The following proposition returns the throughput in the absence of disorientation and misalignment.
\begin{prop}
	In the absence of both disorientation and misalignment, the throughput can be evaluated~as 
	\begin{align}
	\hspace{-0.14cm}	\mathcal{D}^{\text{wo}}\hspace{-0.12cm}\left(r_{\mathrm{th}}\right) &\hspace{-0.12cm}=\hspace{-0.12cm} 
		\left\{ \hspace{-0.19cm}
		\begin{array}{c l}
			r_{\mathrm{th}}\hspace{-0.1cm}\left(\hspace{-0.1cm}1-\frac{\mathrm{G}_{1, 3}^{2, 1}\left( \frac{\Xi^2\frac{2^r_{\mathrm{th}}-2}{\gamma}}{1-\left(\kappa_t^2+\kappa_r^2\right)\left(2^{r_{\mathrm{th}}}-1\right)} \left|\begin{array}{c} 1\\ k_A, m_A, 0 \end{array}\right.\right)}{\Gamma\left(k_A\right)\Gamma\left(m_A\right)}\right), & \\ 
			& \hspace{-5.5cm}\text{ for } r_{\mathrm{th}}< \log_2\left(\frac{1}{\kappa_t^2 + \kappa_r^2}+1\right) \\
			\hspace{-5.5cm} 0, & \hspace{-5.5cm}\text{ for } r_{\mathrm{th}}\geq \log_2\left(\frac{1}{\kappa_t^2 + \kappa_r^2}+1\right)
		\end{array} 
		\right.
		\label{Eq:Thr_wo_final}
	\end{align}
where  \vspace{-0.3cm}
\begin{align}
	k_A = - \frac{b_A}{2 a_A} + \frac{\sqrt{b_A^2 - 4 a_A c_A}}{2 a_A},
	\label{Eq:k_A}
\end{align}  \vspace{-0.5cm}
\begin{align}
	m_A = - \frac{b_A}{2 a_A} - \frac{\sqrt{b_A^2 - 4 a_A c_A}}{2 a_A}
\end{align}
and  \vspace{-0.3cm}
\begin{align}
	\Xi = \sqrt{\frac{k_A m_A}{\Omega_A}}.
	\label{Eq:Xi_A}
\end{align}
In~\eqref{Eq:k_A}--\eqref{Eq:Xi_A},
\begin{align}
	a_A \hspace{-0.15cm}&=\hspace{-0.15cm} \mu_{A}\left(6\right) \mu_{A}\left(2\right) + \left(\mu_{A}\left(2\right)\right)^2 \mu_{A}\left(4\right) - 2\left(\mu_{A}\left(4\right)\right)^2,
	\\
	b_A\hspace{-0.15cm}&=\hspace{-0.15cm} \mu_{A}\left(6\right) \mu_{A}\left(2\right)- 4 \left(\mu_{A}\left(4\right)\right)^2 + 3 \left(\mu_{A}\left(2\right)\right)^2 \mu_{A}\left(4\right),
	\\
	c_A &= 2 \left(\mu_{A}\left(2\right)\right)^2 \mu_{A}\left(4\right)
\end{align}
and \vspace{-0.3cm}
\begin{align}
	\Omega_A = \mu_A\left(2\right),
\end{align}
where \vspace{-0.3cm}
\begin{align}
	\mu_{A}\left(l\right) = \sum_{l_1=0}^{l}&\sum_{l_2=0}^{l_1}\cdots\sum_{l_{N-1}=0}^{l_{N-2}}
	\left(\begin{array}{c}l\\l_1\end{array}\right) \left(\begin{array}{c}l_1\\l_2\end{array}\right)
	\cdots 
	\left(\begin{array}{c}l_{N-2}\\l_{N-1}\end{array}\right)
	\nonumber \\ 
	& \hspace{-0.7cm} \times \mu_{\chi_1}\left(l-l_1\right) \mu_{\chi_2}\left(l_1-l_2\right) 
	\cdots \mu_{\chi_{N-1}}\left(l_{N-1}\right)
	\label{Eq:mu_A}
\end{align}
and \vspace{-0.3cm}
\begin{align}
	\mu_{\chi_i}(l) = \sum_{m=1}^{M}&\sum_{k=1}^{K}
	a_m^{(1)} a_k^{(2)} \left(\frac{c_1}{c_2}\right)^{-\frac{b_m^{(1)}-b_k^{(2)}}{2}} \left(c_1 c_2\right)^{-\frac{b_m^{(1)}+b_k^{(2)}+n}{2}}
	\nonumber \\ & \times
	\Gamma\left(b_{m}^{(1)}+\frac{n}{2}\right) \Gamma\left(b_{k}^{(2)}+\frac{n}{2}\right)
	\label{Eq:mu_x}
\end{align}
\end{prop}
\begin{IEEEproof}
	For brevity, the proof of Proposition 1 is provided in Appendix A. 
\end{IEEEproof}

\textbf{Remark 1.} From~\eqref{Eq:Thr_wo_final}, it becomes apparent that a maximum transmission spectral efficiency threshold exists that is equal to \vspace{-0.2cm}
\begin{align}
	r_{\mathrm{th}}^{m}=2^{\frac{1}{\kappa_t^2 + \kappa_r^2}}-1,
	\label{Eq:gamma_th_m}
\end{align}
beyond which the throughput becomes equal to $0$. Interestingly,~$r_{\mathrm{th}}^{m}$ only depends on the levels of hardware imperfections of the S's transmitter and UAV's receiver.  

\subsection{With disorientation and misalignment} 
The following proposition returns the thoughput in the presence of disorientation and misalignment.
\begin{prop}
	In the presence of disorientation and misalignment, the throughput can be assessed~as in~\eqref{Eq:P_o_w}, given at the top of the next page. 
	\begin{figure*}
	\begin{align}
		\mathcal{D}^{\text{w}}&(r_{\mathrm{th}})=
		\left\{
		\begin{array}{c l}
		 r_{\mathrm{th}}\left(1-\frac{\zeta \mathrm{G}_{5,3}^{1,4}\left(\left.\begin{array}{c}1-k_A, 1-m_A, \frac{1-\zeta}{2}, \frac{2-\zeta}{2},1 \\ 0, \frac{1-\zeta}{2}, - \frac{\zeta}{2}\end{array}\right| \frac{B_o^2}{\Xi^2} \frac{\gamma}{2^r_{\mathrm{th}}-1}\right)}{2\Gamma\left(k_A\right)\Gamma\left(m_A\right)}\right), & \text{ for } r_{\mathrm{th}}< \log_2\left(\frac{1}{\kappa_t^2 + \kappa_r^2}+1\right) \\
		 0, & \text{ for } r_{\mathrm{th}} \geq \log_2\left(\frac{1}{\kappa_t^2 + \kappa_r^2}+1\right)
		\end{array}
		\right.
		\label{Eq:P_o_w}
	\end{align}
\vspace{-0.5cm}
	\hrulefill
	\end{figure*}
\end{prop} 
\begin{IEEEproof}
	For brevity, the proof of proposition 2 is given in Appendix B.  
\end{IEEEproof}

\section{Results \& Discussion}\label{S:Results} 

This section is devoted to verify the theoretical framework that has been presented in Section~\ref{S:E2eChannel}, by means of Monte Carlo simulations and provide an insightful discussion of the effects of fading, disorientation and misalignment in RIS-assisted UAV wireless systems. Along these lines, the following insightful scenario is considered. It is assumed that S-RIS channel coefficients follow independent and identical Nakagami-$m$ distributions with spread parameters $\Omega=1$; thus, $M=1$, $a_{1}^{(1)}=\frac{m^m}{\Gamma(m)}$, $b_1^{(1)}=m$, and $c_1=m$, where $m$ represents the shape parameter. The Rice distribution is used to model the channel coefficients  of the RIS-UAV link. For an accurate approximation of the Rice distribution, we select $K=20$, 
\begin{align}
	a_{k}^{(2)} = \frac{\delta\left(K_r, k\right)}{\sum_{k_1=1}^K \delta\left(K_r, k_1\right) \Gamma\left(b_{k_1}^{(2)}\right) c_2^{-b_{k_1}^{(2)}}},
	\label{Eq:a_k_rice}
\end{align}    
where 
\begin{align}
	\delta\left(K_r, k\right) = \frac{K_r^{k-1}\left(1+K_r\right)^k}{\exp\left(K_1\right) \left( (n-1)!\right)^2},
\end{align}
while $c_2=1+K_r$, and $b_{k}^{(2)}=k$. Moreover, note that in~\eqref{Eq:a_k_rice}, $K_r$ stands for the $K-$factor of the Rice distribution.
Unless otherwise stated, the following scenario is considered. The transmission frequency is set to $100\text{ }\mathrm{GHz}$, $L_1=10\text{ }\mathrm{m}$ and $L_2=5\text{ }\mathrm{m}$. Moreover, $w_o=1\text{ }\mathrm{mm}$, $\theta=-\frac{\pi}{4}$ and $\phi=\frac{4\pi}{3}$, while $\sigma_p=0.05\text{ }\mathrm{rad}$, $\sigma_o=0.1\text{ }\mathrm{rad}$, and $d_x=0.1$. Finally, $N=16$ and $K=5\text{ }\mathrm{dB}$. 

\begin{figure}
	\centering\includegraphics[width=0.9\linewidth,trim=0 0 0 0,clip=false]{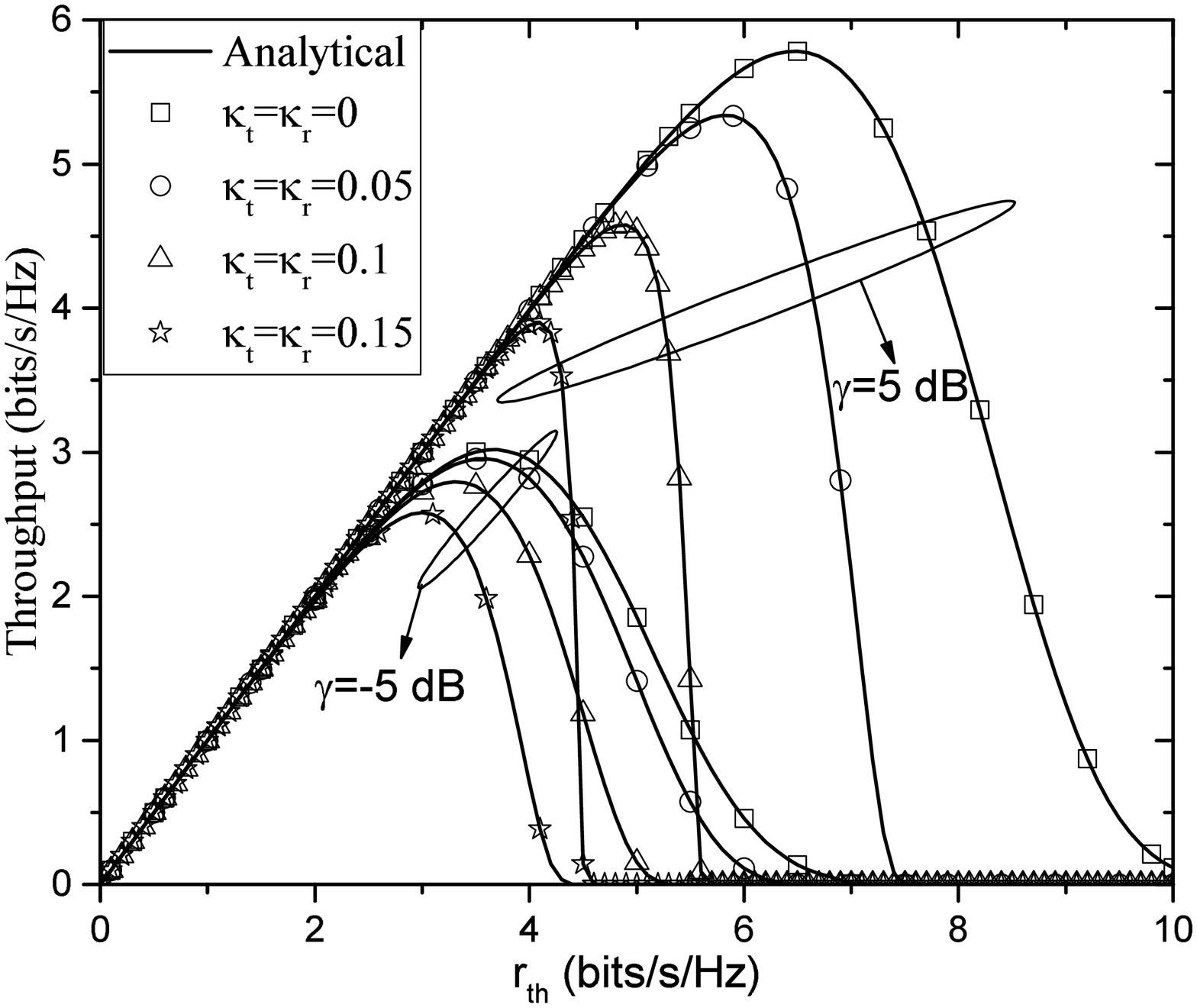}
	\caption{Throughput vs $r_{\mathrm{th}}$, for different values of $\gamma$ and $\kappa_t=\kappa_r$.}
	\label{Fig:D_vs_gth}
\end{figure}

Figure~\ref{Fig:D_vs_gth} quantifies the impact of transceivers hardware imperfections on the throughput performance of the RIS-assisted UAV wireless system. In particular, the achievable throughput is plotted as a function of $r_{\mathrm{th}}$, for different values of $\gamma$ and $\kappa_t=\kappa_r$. From this figure, it becomes evident that for given $\kappa_t=\kappa_r$ and $\gamma$, an optimal transmission scheme spectral efficiency exists, $r_{\mathrm{th}}^{*}$. For $r_{\mathrm{th}}<r_{\mathrm{th}}^{*}$, the achievable throughput increases, as $r_{\mathrm{th}}$ increases. On the other hand, for $r_{\mathrm{th}}>r_{\mathrm{th}}^{*}$, the achievable throughput decreases, as $r_{\mathrm{th}}$ increases. Moreover, for fixed $r_{\mathrm{th}}$ and $\kappa_t=\kappa_r$, as $\gamma$ increases, the achievable throughput also increases. For example, for $r_{\mathrm{th}}=3\text{ }\mathrm{bits/s/Hz}$ and $\kappa_t=\kappa_r=0.1$, the achievable throughput increases from $2.72$ to $3\text{ }\mathrm{bits/s/Hz}$, as $\gamma$ increases from $-5$ to $5\text{ }\mathrm{dB}$. Finally, for given $\gamma$ and $r_{\mathrm{th}}$, as the level of transceiver hardware imperfections increases, an achievable throughput degradation is observed.        

\begin{figure}
	\centering\includegraphics[width=0.9\linewidth,trim=0 0 0 0,clip=false]{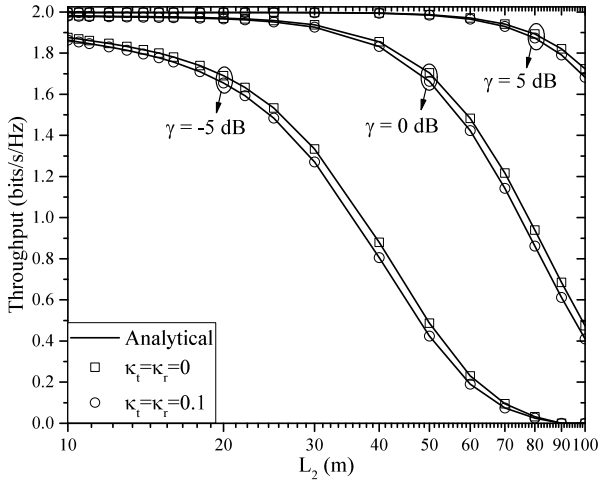}
	\caption{Throughput vs $L_2$, for different values of $\gamma$ and $\kappa_t=\kappa_r$.}
	\label{Fig:D_vs_L2}
\end{figure}

Figure~\ref{Fig:D_vs_L2} depicts the achievable throughput as a function of $L_2$, for different values of $\gamma$ and $\kappa_t=\kappa_r$, assuming $r_{\mathrm{th}}=2\text{ }\mathrm{bits/s/Hz}$. As expected, for given $\kappa_t=\kappa_r$ and $\gamma$, as $L_2$ increases, the impact of disorientation becomes more severe; thus, the throughput decreases. For example, for  $\kappa_t=\kappa_r=0$, i.e., the ideal RF front-end case, and $\gamma=-5\text{ }\mathrm{dB}$, the throughput decreases from $1.9$ to $0\text{ }\mathrm{bits/s/Hz}$, as $L_2$ increases from $10$ to $100\text{ }\mathrm{m}$. Moreover, for fixed $L_2$ and $\gamma$, as the level of transceiver hardware imperfections increases, the achievable throughput decreases. For instance, for $L_2=50\text{ }\mathrm{m}$ and $\gamma=-5\text{ }\mathrm{dB}$, the achievable throughput changes from $0.49$ to $0.42\text{ }\mathrm{bits/s/Hz}$, as $\kappa_t=\kappa_r$ increases from $0$ to $0.1$. Finally, for given $L_2$ and $\kappa_t=\kappa_r$, the achievable throughput increases, as $\gamma$ increases.

\section{Conclusions}\label{S:Conclusions} 

This paper was devoted to analyze the throughput performance of RIS-assisted UAV wireless systems in the absence and presence of stochastic beam misalignment and disorientation as well as transceiver hardware imperfections. In this direction, novel closed-form expressions  for the achievable throughput for both cases of presence and absence of stochastic beam misalignment and disorientation were derived. Our results highlighted the importance of accurate modeling the aforementioned phenomena as well as small-scale fading, when assessing the throughput performance of RIS-assisted UAV wireless systems. Finally, it became evident that an optimal transmission spectral efficiency exists for which the achievable throughput is maximized.        

\section*{Appendices} 

\section*{Appendix A}
\section*{Proof of Proposition 1}

The throughput is defined~as 
\begin{align}
	\mathcal{D}^{\text{wo}}\left(r_{\mathrm{th}}\right) = r_{\mathrm{th}} \Pr\left(r>r_{\mathrm{th}}\right),
	\label{Eq:Thr_def}
\end{align}
where
\begin{align}
	r = \log_2\left(1+\gamma_u\right).
	\label{Eq:r_appA}
\end{align}
In the absence of disorientation and misalignment,~\eqref{Eq:r_appA} can be rewritten~as
\begin{align}
	r = \log_2\left(1+ \frac{A^2}{\left(\kappa_t^2+\kappa_r^2\right)A^2+\frac{1}{\tilde{\gamma}}}\right).
	\label{Eq:r_s2_appA}
\end{align}  
By applying~\eqref{Eq:r_s2_appA} into~\eqref{Eq:Thr_def}, the throughput can be rewritten~as
\begin{align}
	\mathcal{D}^{\text{wo}}\left(r_{\mathrm{th}}\right) = r_{\mathrm{th}} \Pr\left(\log_2\left(1+ \frac{A^2}{\left(\kappa_t^2+\kappa_r^2\right)A^2+\frac{1}{\tilde{\gamma}}}\right)>r_{\mathrm{th}}\right),
	\label{Eq:Thr_s2_def}
\end{align}
or equivalently
\begin{align}
	&\mathcal{D}^{\text{wo}}\left(r_{\mathrm{th}}\right) \hspace{-0.1cm} = \hspace{-0.1cm} r_{\mathrm{th}}
	\nonumber \\ &\times \Pr\left( A^2 \left(1-\left(2^{r_{\mathrm{th}}}-1\right)\left(\kappa_t^2+\kappa_r^2\right) \right)>\frac{2^{r_{\mathrm{th}}}-1}{\tilde{\gamma}} \right)
\end{align}
or
\begin{align}
	&\mathcal{D}^{\text{wo}}\left(r_{\mathrm{th}}\right)\hspace{-0.1cm} = \hspace{-0.1cm} r_{\mathrm{th}}
	\nonumber \\ & \times \left(1-\Pr\left( A^2 \left(1-\left(2^{r_{\mathrm{th}}}-1\right)\left(\kappa_t^2+\kappa_r^2\right) \right)\leq\frac{2^{r_{\mathrm{th}}}-1}{\tilde{\gamma}} \right)\right).
	\label{Eq:Thr_s3_def}
\end{align}
For $\left(2^{r_{\mathrm{th}}}-1\right)\left(\kappa_t^2+\kappa_r^2\right)>1$, $A^2 \left(1-\left(2^{r_{\mathrm{th}}}-1\right)\left(\kappa_t^2+\kappa_r^2\right) \right)$ is always negative, thus, $\Pr\left( A^2 \left(1-\left(2^{r_{\mathrm{th}}}-1\right)\left(\kappa_t^2+\kappa_r^2\right) \right)\leq\frac{2^{r_{\mathrm{th}}}-1}{\tilde{\gamma}} \right)=1$, and
 \begin{align}
 	&\mathcal{D}^{\text{wo}}\left(r_{\mathrm{th}}\right) = 0, \text{ for } r_{\mathrm{th}}> \log_2\left(1+\frac{1}{\kappa_t^2+\kappa_r^2} \right).
 	\label{Eq:Thr_0}
\end{align}
On the other hand, for $r_{\mathrm{th}}< \log_2\left(1+\frac{1}{\kappa_t^2+\kappa_r^2} \right)$,~\eqref{Eq:Thr_s3_def} can be rewritten~as
\begin{align}
		&\mathcal{D}^{\text{wo}}\left(r_{\mathrm{th}}\right) = r_{\mathrm{th}}
		\nonumber \\ & \times \left(1-F_A\left(\sqrt{\frac{1}{\left(1-\left(2^{r_{\mathrm{th}}}-1\right)\left(\kappa_t^2+\kappa_r^2\right) \right)} \frac{2^{r_{\mathrm{th}}}-1}{\tilde{\gamma}}} \right) \right),
		\label{Eq:Thr_s4_def}
\end{align}
where $F_{A}\left(\cdot\right)$ stands for the CDF of $A$. 
From~\eqref{Eq:Thr_s4_def}, it becomes evident that in order to evaluate the achievable throughput, we need first to extract the CDF of $A$. Towards this direction, from~\eqref{Eq:A_final}, we can rewrite $A$~as 
\begin{align}
	A = \sum_{i=1}^{N}\chi_i,
	\label{Eq:A_final_2}
\end{align}
where 
\begin{align}
	\chi_i=\left|h_i\right| \left|g_i\right|.
	\label{Eq:chi_i}
\end{align}
Since $\left|h_i\right|$ and $\left|g_i\right|$, the PDF of $\chi_i$ can be analytically evaluated~as 
\begin{align}
	f_{\chi_i}(x) = \int_{0}^{\infty} \frac{1}{y} f_{h_i}(y) f_{g_i}\left(\frac{x}{y}\right)\,\mathrm{d}y,
\end{align}
which, by applying~\eqref{Eq:f_h_i} and~\eqref{Eq:f_g_i} can be equivalently written~as 
\begin{align}
	f_{\chi_i}(x) = \sum_{m=1}^{M} \sum_{k=1}^{K} 4 a_{m}^{(1)} a_k^{(2)} x^{2 b_{k}^{(2)}-1} \mathcal{I}_{m,k}\left(x\right),
	\label{Eq:f_chi_i}
\end{align}
where 
\begin{align}
	\mathcal{I}_{m,k}\hspace{-0.1cm}\left(x\right) \hspace{-0.1cm} = \hspace{-0.1cm} \int_0^{\infty} \hspace{-0.1cm} y^{2 b_m^{(1)} - 2 b_k^{(2)}-1} \exp\left(-c_1 y^2 - \frac{c_2 x^2}{y^2}\right)\,\mathrm{d}y. 
	\label{Eq:I}
\end{align}
With the aid of~\cite[Eq. (3.478/4)]{B:Gra_Ryz_Book},~\eqref{Eq:I} can be expressed in closed-form~as 
\begin{align}
	\mathcal{I}_{m,k}\hspace{-0.1cm}\left(x\right)\hspace{-0.1cm} =\hspace{-0.1cm} \left(\frac{c_1}{c_2}\right)^{-\frac{ b_m^{(1)}- b_k^{(2)}}{2}} x^{{b_m^{(1)}}-b_k^{(2)}} \mathrm{K}_{{b_m^{(1)}}-{b_k^{(2)}}}\left(2\sqrt{c_1 c_2} x\right).
	\label{Eq:I_s2}
\end{align}
By~applying~\eqref{Eq:I_s2} into~\eqref{Eq:f_chi_i}, we get 
\begin{align}
	f_{\chi_i}(x) = \sum_{m=1}^{M}  \sum_{k=1}^{K} & 4 a_{m}^{(1)} a_k^{(2)} \left(\frac{c_1}{c_2}\right)^{-\frac{ b_m^{(1)}- b_n^{(2)}}{2}} x^{{b_m^{(1)}}+b_k^{(2)}-1}
	\nonumber \\ & \times
	\mathrm{K}_{{b_m^{(1)}}-{b_k^{(2)}}}\left(2\sqrt{c_1 c_2} x\right),
	\label{Eq:f_chi_i_semifinal}
\end{align}
or  
\begin{align}
	f_{\chi_i}(x) &= \sum_{m=1}^{M}   \sum_{k=1}^{K} a_m^{(1)} a_k^{(2)}
	\left(\frac{c_1}{c_2}\right)^{-\frac{b_m^{(1)}-b_k^{(2)}}{2}} \left(c_1 c_2\right)^{-\frac{b_m^{(1)}+b_k^{(2)}}{2}}
	\nonumber \\ & \times 
	\Gamma\left({b_m^{(1)}}\right)
	\Gamma\left(b_k^{(2)}\right) 
	f_{\mathrm{K}_G}^{(m,k)}\left(x\right),
	\label{Eq:f_chi_i_final}
\end{align}
where \vspace{-0.2cm}
\begin{align}
	f_{\mathrm{K}_G}^{(m,k)}\left(x\right) &= 
	\frac{4\left(c_1 c_2\right)^{\frac{b_m^{(1)}+b_k^{(2)}}{2}}x^{b_m^{(1)}+b_k^{(2)}-1}\mathrm{K}_{b_m^{(1)}-b_k^{(2)}}\left(2\sqrt{c_1 c_2} x\right)}{\Gamma\left(b_m^{(1)}\right)\Gamma\left(b_k^{(2)}\right)} 
	\label{Eq:f_K_G}
\end{align}
Notice that~\eqref{Eq:f_K_G} is a special case of the PDF of the generalized-K distribution. Thus, from~\eqref{Eq:f_chi_i_final}, it becomes apparent that $\chi_i$ follows a mixture generalized-K distribution. Likewise, from~\eqref{Eq:A_final_2}, we observe that $A$ is a sum of $N$ mixture generalized-K distributed RVs; hence, according to~\cite[Eq. (7)]{Peppas2011}, its  CDF can be obtained as \begin{align}
	F_{A}(x) = \frac{1}{\Gamma\left(k_A\right)\Gamma\left(m_A\right)} \mathrm{G}_{1, 3}^{2, 1}\left(\Xi^2 x^2\left|\begin{array}{c} 1\\ k_A, m_A, 0 \end{array}\right.\right),
	\label{Eq:F_A}
\end{align}
where~\eqref{Eq:k_A}--\eqref{Eq:mu_A} can be extracted by applying~\cite[Eq. (17)]{Peppas2011}. Finally, $\mu_{\chi_i}(l)$ can be analytically evaluated~as \vspace{-0.1cm}
\begin{align}
	\mu_{\chi_i}(l) = \int_0^{\infty} x^{n} f_{\chi_i}(x)\,\mathrm{d}x,
\end{align}
which, by applying~\eqref{Eq:f_chi_i_semifinal}, can be rewritten~as \vspace{-0.1cm}
\begin{align}
	\mu_{\chi_i}(l) = \sum_{m=1}^{M}  \sum_{k=1}^{K} & 4 a_{m}^{(1)} a_k^{(2)} \left(\frac{c_1}{c_2}\right)^{-\frac{ b_m^{(1)}- b_n^{(2)}}{2}} \mathcal{K}_{m,n}, 
	\label{Eq:mu_chi_i_l_s0}
\end{align}
where \vspace{-0.2cm}
\begin{align}
	\mathcal{K}_{m,n} = \int_0^{\infty}   x^{{b_m^{(1)}}+b_k^{(2)}+n-1}
	\mathrm{K}_{{b_m^{(1)}}-{b_k^{(2)}}}\left(2\sqrt{c_1 c_2} x\right)\,\mathrm{d}x.
	\label{Eq:K}
\end{align}
By using~\cite[Eq. (6.561/16)]{B:Gra_Ryz_Book} in~\eqref{Eq:K}, we obtain 
\begin{align}
	\mathcal{K}_{m,n} = \frac{1}{4} \left(c_1 c_2\right)^{-\frac{{b_m^{(1)}}+b_k^{(2)}+n}{2}}\,\Gamma\left({b_m^{(1)}}+\frac{n}{2} \right)\,\Gamma\left({b_k^{(2)}}+\frac{n}{2} \right).
	\label{Eq:K_s2}
\end{align}
Finally, with the of~\eqref{Eq:K_s2}, \eqref{Eq:mu_chi_i_l_s0} can be written as~\eqref{Eq:mu_x}.
Next, by applying~\eqref{Eq:F_A} into~\eqref{Eq:Thr_s4_def} and accounting~\eqref{Eq:Thr_0}, we obtain~\eqref{Eq:Thr_wo_final}. This concludes the proof.  

\section*{Appendix B}
\section*{Proof of Proposition 2}

In the case the system experiences disorientation and misalignment, by following the same steps as in the Proof of Proposition 1, in the case $\left(2^{r_{\mathrm{th}}}-1\right)\left(\kappa_t^2+\kappa_r^2\right)\leq 1$ the throughput can be expressed~as
\begin{align}
		&\mathcal{D}^{\text{wo}}\left(r_{\mathrm{th}}\right) = 0,\, \text{for}\, r_{\mathrm{th}}\geq \log_2\left(1+\frac{1}{\kappa_t^2+\kappa_r^2}\right).
		\label{Eq:D1}
\end{align}
On the other hand, for $r_{\mathrm{th}}< \log_2\left(1+\frac{1}{\kappa_t^2+\kappa_r^2}\right)$
\begin{align}
	&\mathcal{D}^{\text{wo}}\left(r_{\mathrm{th}}\right) = r_{\mathrm{th}}
	\nonumber \\ & \times \left(1-F_{A_{e2e}}\left(\sqrt{\frac{1}{\left(1-\left(2^{r_{\mathrm{th}}}-1\right)\left(\kappa_t^2+\kappa_r^2\right) \right)} \frac{2^{r_{\mathrm{th}}}-1}{\tilde{\gamma}}} \right) \right),
	\label{Eq:Thr_e2e_s1_def}
\end{align}
where $F_{A_{e2e}}\left(\cdot\right)$ stands for the CDF of $A_{e2e}$.  

To evaluate~\eqref{Eq:Thr_e2e_s1_def}, we first need to find the distribution of $A_{e2e}$. Since $h_g$ and $A$ are independent RVs, the CDF of $A_{e2e}$ can be written~as 
\begin{align}
	F_{A_{e2e}}(x) = \int_{0}^{B_o} F_{A}\left(\frac{x}{y}\right) f_{h_g}(y)\,\mathrm{d}y. 
	\label{Eq:F_A_e2e}
\end{align}
By employing~\eqref{Eq:f_h_g} and~\eqref{Eq:F_A},~\eqref{Eq:F_A_e2e} can be expressed~as 
\begin{align}
	F_{A_{e2e}}(x) =  \frac{\rho}{B_o^{\zeta}}\frac{1}{\Gamma\left(k_A\right)\Gamma\left(m_A\right)} \mathcal{J}(x),
	\label{Eq:F_A_e2e_appendix}
\end{align}
where 
\begin{align}
	\mathcal{J}(x) = \int_{0}^{B_o} y^{\zeta-1} \mathrm{G}_{1, 3}^{2, 1}\left(\Xi^2 \frac{x^2}{y^2}\left| \begin{array}{c} 1 \\ k_A, m_A, 0 \end{array} \right. \right) \, \mathrm{d}y. 
	\label{Eq:J}
\end{align}
With the aid of~\cite[Eq. (07.34.17.0012.01)]{WS:mathematica_function},~\eqref{Eq:J} can be equivalently written as 
\begin{align}
	\mathcal{J}(x) = \int_{0}^{B_o} y^{\zeta-1} \mathrm{G}_{ 3, 1}^{ 1, 2}\left(\Xi^{-2} \frac{y^2}{x^2}\left| \begin{array}{c}  1-k_A, 1-m_A, 1 \\ 0 \end{array} \right. \right) \, \mathrm{d}y,
	\label{Eq:Js2}
\end{align}
which, by employing~\cite{W:07.34.21.0084.01}, yields 
\begin{align}
	\mathcal{J}(x) = \frac{B_o^{\zeta}}{2\zeta} \mathrm{G}_{5,3}^{1,4}\left(\left.\begin{array}{c}1-k_A, 1-m_A, \frac{1-\zeta}{2}, \frac{2-\zeta}{2},1 \\ 0, \frac{1-\zeta}{2}, - \frac{\zeta}{2}\end{array}\right| \frac{B_o^2}{\Xi^2 x^2}\right).
	\label{Eq:Js3}
\end{align}
With the aid of~\eqref{Eq:Js3},~\eqref{Eq:F_A_e2e_appendix} can be written~as
\begin{align}
	F_{A_{e2e}}&(x)= \frac{\zeta}{2\Gamma\left(k_A\right)\Gamma\left(m_A\right)}
	\nonumber \\ & \hspace{-0.6cm}\times \mathrm{G}_{5,3}^{1,4}\left(\left.\begin{array}{c}1-k_A, 1-m_A, \frac{1-\zeta}{2}, \frac{2-\zeta}{2},1 \\ 0, \frac{1-\zeta}{2}, - \frac{\zeta}{2}\end{array}\right| \frac{B_o^2}{\Xi^2 x^2}\right).
	\label{Eq:F_A_e2e_final_general}
\end{align} 
By applying~\eqref{Eq:F_A_e2e_final_general} into~\eqref{Eq:Thr_e2e_s1_def} and combining the resulting formula with~\eqref{Eq:D1}, we obtain~\eqref{Eq:P_o_w}. This concludes the proof. 
 
\balance
\bibliographystyle{IEEEtran}
\bibliography{IEEEabrv,References}

\end{document}